\documentclass[paper]{JHEP3} % 10pt is ignored!

\usepackage{epsfig,multicol,bbm}

\newcommand\fverb{\setbox\fverbbox=\hbox\bgroup\verb}
\newcommand\fverbdo{\egroup\medskip\noindent%
			\fbox{\unhbox\fverbbox}\ }
\newcommand\fverbit{\egroup\item[\fbox{\unhbox\fverbbox}]}
\newbox\fverbbox

\newcommand{\be}{\begin{eqnarray}}
\newcommand{\ee}{\end{eqnarray}}

\title{Large scale directional anomalies in the WMAP 5yr ILC map}

\author{Alessandro Gruppuso
\\
	INAF/IASF-BO, Istituto di Astrofisica Spaziale e Fisica Cosmica di Bologna,\\
	via Gobetti 101, I-40129 Bologna, Italy
\\
	INFN, Sezione di Bologna,
	Via Irnerio 46, I-40126 Bologna, Italy
\\
	E-mail: \email{gruppuso@iasfbo.inaf.it}}

\author{Krzysztof~M.~Gorski
\\
Jet Propulsion Laboratory, California Institute of Technology, 
4800 Oak Grove Drive, Pasadena CA 91109, U. S. A.
\\
	Warsaw University Observatory, Aleje Ujazdowskie 4, 00-478 
Warszawa, Poland
\\
	E-mail: \email{krzysztof.m.gorski@jpl.nasa.gov}}

\received{\today} 		%%
\accepted{\today}		%% These are for published papers.

\abstract{We study the alignments of the low multipoles of CMB anisotropies with specific directions in the sky 
(i.e.~the dipole, the north Ecliptic pole, the north Galactic pole and the north Super Galactic pole).
Performing $10^5$ random extractions we have found that: 1) separately quadrupole and octupole are mildly orthogonal to the dipole 
but when they are considered together, in analogy to \cite{Copi2006}, we find an unlikely orthogonality at the level of $0.8\%$ C.L.;
2)  the multipole vectors associated to $\ell=4$ are unlikely aligned with the dipole at $99.1 \%$ C.L.; 
3) the multipole vectors associated to $\ell=5$ are mildly orthogonal to the dipole but when we consider only maps that show exactly the same correlation 
among the multipoles as in the observed WMAP 5yr ILC, these multipole vectors are unlikely orthogonal to the dipole at $99.7 \%$ C.L..}


\begin{document} 

%\maketitle  IS IGNORED %%%%%%%%%%%

\section{Introduction}
\label{intro}
The anisotropy pattern of the cosmic microwave background
(CMB), obtained by Wilkinson Microwave Anisotropy Probe
(WMAP), probes cosmological models with unprecedented precision (see \cite{WMAP5yr}
and references therein). WMAP data are largely consistent
with the concordance $\Lambda$ cold dark matter ($\Lambda$CDM) model, 
but there
are some interesting deviations from it, in particular on the largest
angular scales. 
%They can be divided in three categories.
%{\it 1) Lack of power at large scales}.
%The angular correlation function is found to be uncorreleted (i.e. consistent with $0$) for angles larger than $60$ degrees. In \cite{Copi2008, Copi2006} 
%it has been shown that the probability associated to this event is low as $0.15 \%$. 
%Still in this category we mention the surprisingly low amplitude of the quadrupole term of the angular power spectrum (APS), already found by Cosmic Background Explorer (COBE) \cite{Smoot1992,Hinshaw1996}, and now confirmed by
%WMAP \cite{WMAP5yr}. 
%{\it 2) Unlikely alignments of low multipoles}.
Among these deviations, we focus on the alignments of low multipoles. 
Unlikely (for a statistically isotropic random field) alignment of the quadrupole and the octupole is
described in references
\cite{copi2004,weeks,lowalignments,CopiMNRAS2006,Gruppuso:2009ee}. 
Moreover, both quadrupole and octupole align with the CMB dipole
\cite{CopiMNRAS2006,Copi2006}. 
Other unlikely alignments are described in \cite{abramo}. 
%{\it 3) Hemispherical asymmetries}.
%It is found that the power coming separately from the two hemispheres (defined by the ecliptic plane) 
%is too asymmetric (especially at low $\ell$) \cite{Eriksen}.

We study directionality anomalies considering estimators defined for each multipole $\ell$.
In the present paper we limit our analysis to the range $\Delta \ell=2-7$.
We take into account two different Monte-Carlo (henceforth MC) simulations.
In the first one, we study directional anomalies considering the coefficients of the spherical harmonics expansions
$a_{\ell m}$ randomly extracted from a Gaussian distribution.
In the second one we consider maps in which the relative alignments of the low multipoles
are fixed and given by the WMAP 5yr ILC map (henceforth called ILC map for sake of brevity).
Both MCs are performed taking into account WMAPÕs anisotropic pixel noise (see below for details).
Through the first MC, we obtain that quadrupole and octupole are mildly unlikely orthogonal to the dipole (this anomaly become 
stronger when they are combined in analogy to what has been performed in \cite{Copi2006}). 
Moreover we find that $\ell=4$ is unlikely aligned with the dipole and $\ell=5$ is very mildly unlikely orthogonal to the dipole. 
Through the second MC, we find that quadrupole and octupole are (mildly) unlikely orthogonal to the dipole
and combining them together (still in analogy to \cite{Copi2006}) we do not find any increase of this anomaly.
In this case $\ell=4$ is unlikely aligned with the dipole and $\ell=5$ is unlikely orthogonal to the dipole. 
We do not obtain unlikely alignments with the dipole for $\ell=6$ and $7$ for both MCs.

The paper is organized as follows: in Section \ref{estimators} we present the Multipole Vectors expansion and the tools
that have been used to perform the statistical analysis; such analysis is given in Section \ref{analysis} where we present 
also our main results; conclusions are drawn in Section \ref{conclusions}.
%Some comments about the shape of the distribution of the alignment estimators can be found in Appendix \ref{shape} 
%whereas some technicalities about the performed simulations are described in Appendix \ref{technicalities}.  
Some technicalities about the performed simulations can be found in Appendix \ref{appendix}.  

\section{Tools and estimators}
\label{estimators}
The alignment of multipoles can be defined
using a new representation of CMB anisotropy maps
where the $a_{\ell m}$ 
(coefficients of the expansion over the basis of spherical harmonics)
are replaced by vectors \cite{copi2004,weeks}.
In particular, each multipole order $\ell$ is represented by $\ell$
unit vectors $\hat u_{i}$ and one amplitude $A$
\be
 a_{\ell m} \leftrightarrow A^{(\ell)}, \hat u_1, \, \cdot \cdot \cdot \, , \hat u_{\ell}
\, .
\label{association}
\ee
Note that the number of independent objects is the same in the l.h.s and r.h.s. of equation (\ref{association}):
$2 \ell +1$ for $a_{\ell m}$ 
equals $3 \ell$ (numbers of components of the vectors) $+1$ (given by $A^{(\ell)}$) $-\ell$ (because there are $\ell$ constraints due to the normalization conditions of the vectors). 
%One of the advantage of Multipole Vectors representation is that from 
%these unit vectors one can easily construct scalar quantities that 
%are invariant under rotation. 
%Note that is not equally easy to obtain scalar quantities directly 
%from the $a_{\ell m}$ coefficients since they depend on the 
%coordinate system. 
For a more detailed explanation of equation (\ref{association})
and of the properties of that association
see for example references \cite{copi2004,weeks, abramo, DSCgruppuso}.

Unfortunately, an explicit analytical expression for the association
given in equation (\ref{association}), is possible only for $\ell =1$
and for $\ell \neq 1$ numerical method are needed\footnote{Indeed, for $\ell=2$ it is possible to obtain the multipole vectors
computing the eigenvectors of a symmetric and traceless tensor representing the quadrupole, see \cite{landmagueijo,dennis}.}.
%For $\ell =1$ we have the 
%following analytical solution
%\begin{eqnarray}
%d_x &=& \mp a_{1 1}^{(R)}/\sqrt{a_{1 0}^2/2+((a_{1 1}^{(R)})^2+(a_{1 1}^{(I)})^2)}  \, , \label{dx}\\
%d_y &=& \pm a_{1 1}^{(I)}/\sqrt{a_{1 0}^2/2+((a_{1 1}^{(R)})^2+(a_{1 1}^{(I)})^2)} \, , \label{dy}\\
%d_z &=& \pm a_{1 0}/\sqrt{a_{1 0}^2+2 ((a_{1 1}^{(R)})^2+(a_{1 1}^{(I)})^2)} \, , \label{dz}\\
%A^{(1)} &=& \mp {1\over 2} \sqrt{3\over \pi} 
%\sqrt{a_{1 0}^2+2((a_{1 1}^{(R)})^2+(a_{1 1}^{(I)})^2)} \, ,
%\label{A1}
%\end{eqnarray}
%where $\vec d = (d_x, d_y,d_z )$ and the labels $(R)$ and $(I)$ stand for real and imaginary part.
%
The Copi et al.'s algorithm 
%\footnote{For sake of brevity, we call the routine that performs the Multipole Vectors decomposition ``Copi's routine'' or ``Copi's algorithm''. See footnote 5 or Ref.~\cite{copi2004} to give credit to all the developers of that code.}
(which use is acknowledged here) for constructing multipole vectors from a standard spherical harmonic decomposition is described in \cite{copi2004} 
and the implementation of it is publicly available\footnote{http://www.phys.cwru.edu/projects/mpvectors/}. 
Other methods exist but, as far as we know, their implementation is not publicly available
on a standard platform (see for example \cite{weeks,Katz} 
where the problem of finding $\ell$ vectors is translated into the problem of finding
the zeros of a polynomial of degree $2\ell$).

In order to investigate whether a map presents unlikely directions with respect to random 
%Gaussian 
CMB sky 
extractions\footnote{See Section \ref{statistics} for a detailed description of the performed extractions.}, we define the following
estimators (scalars quantities) for each multipole order $\ell$:
\be
S_{\ell 1} = \sum_{i=1}^{\ell} \,  | \hat u_{i} \cdot \hat d | / \ell  \, ,
\label{estimatorsdefinition}
\ee
where $\hat u_{i}$ represent the multipole vectors associated to a multipole order $\ell$ of a map, $\hat d$ is a fixed direction in the sky and 
where the absolute value is taken into account since multipole vectors are vectors without ``head'' (this is the so called reflection symmetry).
The division by $\ell$ is considered in order to define the estimators (\ref{estimatorsdefinition}) between $0$ and $1$. 

%\subsection{Copi et al.'s Estimator}
%\label{Copi et al.'s Estimator}

For sake of comparison, in this paragraph we present the estimator used in \cite{CopiMNRAS2006,Copi2006}
(where it is shown that quadrupole and octupole are unlikely aligned with the dipole).
The definition of that estimator is the following
%This has been demonstrated using the following estimator
\be
S_{321} = {1\over 4} \left( |\vec q \cdot \hat d| + \sum_{i=1}^3 |\vec o_i \cdot \hat d | \right)
\, ,
\label{copiestimator}
\ee
where $\vec q$ and $\vec o_i$ are the area vectors built from the quadrupole and octupole multipole vectors respectively.
More explicitly: 
\be
\vec q = \hat q_{21} \times \hat q_{22}
\, ,
\ee
and 
\be
\vec o_1 &=& \hat o_{32} \times \hat o_{33} \, , \\
\vec o_2 &=& \hat o_{33} \times \hat o_{31} \, , \\
\vec o_3 &=& \hat o_{31} \times \hat o_{32}
\, ,
\ee
where $\hat q_{2i}$ with $i=1,2$ are the two multipole vectors associated to the quadrupole
and $\hat o_{3j}$ with $j=1, 2, 3$ are the three multipole vectors associated to the octupole.
In eq.~(\ref{copiestimator}) the division by $4$ has been taken into account to define the estimator 
$S_{321}$ in the range  $[0,1]$ and the absolute values are considered to fulfill the reflection symmetry. 

\section{The Analysis}
\label{analysis}

\subsection{Description of the Random Extractions}
\label{statistics}

As already stated in Section \ref{intro} we perform two kinds of MCs in order to study how likely
the ILC map \footnote{The 5 years WMAP ILC map, as well as other CMB data products, is publicly available at
the Lambda web site: http://lambda.gsfc.nasa.gov/.} is aligned to some specific direction.

For the first MC we extract $10^5$ random maps from a Gaussian distribution
from an angular power spectrum corresponding to the best fit of WMAP 5yr (see footnote $4$).
For each random extraction, the estimators  $S_{\ell 1}$  (with $\ell = 2, 3, 4, 5, 6, 7$) and $S_{321}$ are computed 
for some direction of interest such as the Dipole (DIP), the north Ecliptic pole (NEP), the north Galactic pole (NGP) and the north Super Galactic pole (NSGP).

%Given the ILC map\footnote{The 5 years WMAP ILC map is publicly available at
%the Lambda web site: http://lambda.gsfc.nasa.gov/.} 
%we study how likely it is aligned to some specific direction.

For the second MC, as done in \cite{CopiMNRAS2006}, we freeze the relative alignments of the low multipoles such 
as those that are observed in the ILC map
and perform $10^5$ random rotations of that map. For each random rotation the estimators $S_{\ell 1}$ 
(with $\ell = 2, 3, 4, 5, 6, 7$) and $S_{321}$ are computed for the aforementioned directions (DIP, NEP, NGP and NSGP namely).

Both MCs are performed taking into account WMAP's anisotropic pixel 
noise\footnote{We consider the WMAP V band noise and hits files available at Lambda web site. See footnote 4.}. 
%In all the plots we show the probability distribution function (pdf) for
%each $S_{\ell 1}$ and the vertical line representing the $S_{\ell 1}$ value of the ILC map (henceforth ILC value).
The resolution that has been considered for the analysis is given by the HEALPIx\footnote{http://healpix.jpl.nasa.gov} \cite{gorski} parameter $N_{side}=16$ that 
corresponds to $3072$ pixels\footnote{The Healpix parameter $N_{side}$ is related to the number of pixels $N_{pix}$ through $N_{pix}=12 \, N_{side}^2$.}. 
See appendix \ref{technicalities} for details about the degradation issue.

\subsection{Results}
\label{results}

In this Section we provide the results for both MCs.

Results for the first MC are summarized in Table \ref{tableprobabilities}
where the probabilities to obtain a smaller value for the analyzed directions
are listed for the ILC map.

\begin{table}[ht]
\caption{First MC: Probabilities (in percentage) to obtain a smaller value for the considered directions and estimators} % title of Table
\centering % used for centering table
\begin{tabular}{c c c c c} % centered columns (5 columns)
\hline\hline %inserts double horizontal lines
Estimator &  DIP &  NEP &  NGP  &  NSGP \\ [0.5ex] % inserts table
%heading
\hline % inserts single horizontal line
$S_{21}$ & 6.7 & 42.3 & 9.6 & 41.5 \\ % inserting body of the table
$S_{31}$ & 3.6 & 86.0 & 7.2 & 56.2 \\
$S_{321}^{new}$ & {\bf 0.8} & 73.6 & 2.6 & 49.4 \\
$S_{41}$ & {\bf 99.1} & 2.6 & 48.2 & 74.5 \\
$S_{51}$ & 9.9 & 30.9 & 56.5 & 87.7 \\
$S_{61}$ & 76.2 & 58.8 & 37.5 & 95.0 \\
$S_{71}$ & 54.6 & 4.0 & 41.4 & 88.6 \\  [1ex] % [1ex] adds vertical space
\hline %inserts single line
\end{tabular}
\label{tableprobabilities} % is used to refer this table in the text
\end{table}

Table \ref{tableprobabilities} shows that multipole vectors associated to quadrupole, 
octupole and $\ell=5$ are mildly unlikely orthogonal to the dipole. 
Multipole vectors associated to $\ell=4$ are unlikely aligned to the dipole at the $99.1 \%$ of C.L..

In analogy to what performed in \cite{CopiMNRAS2006,Copi2006}, when $S_{21}$ and $S_{31}$ are combined as follows
\be
S_{321}^{new} = (2 \, S_{21} + 3 \, S_{31} ) /5
\label{newS321eq}
\ee
we find that the percentages to obtain a smaller value for the considered directions are
$0.8$, $73.6$, $2.6$, $49.4$ for DIP, NEP, NGP and NSGP respectively.
This confirms that when quadrupole and octupole are considered together as in eq.~(\ref{newS321eq})
then the anomaly related to the dipole direction becomes more severe. Note that the estimator defined in eq.~(\ref{newS321eq})
differs from eq.~(\ref{copiestimator}) even if both are of course related in information content.
Considering eq.~(\ref{copiestimator}), it is shown in \cite{Copi2006} that the anomaly is as low as $0.3\%$.
Therefore both estimators (i.e. $S_{321}^{new}$ and $S_{321}$) provide the same level of anomaly\footnote{We should 
remind that the ILC map used in \cite{Copi2006} was the WMAP 3 yr ILC map and not the last one available now (analyzed in this paper).}.

\begin{figure}
\includegraphics[width=8.0cm]{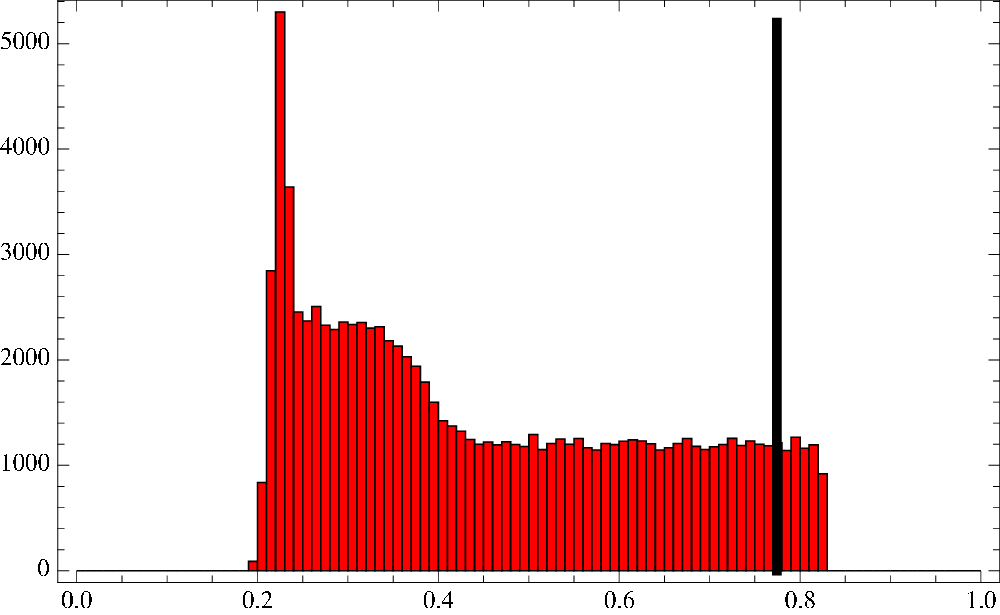}
\caption{Copi et al.'s estimator \cite{Copi2006}, called $S_{321}$ in this paper. 
The black vertical line represent the Dipole direction. The plot presents the counts (y-axis) versus the statistic (x-axis).}
\label{S231}
\end{figure}

\begin{figure}
\includegraphics[width=8.0cm]{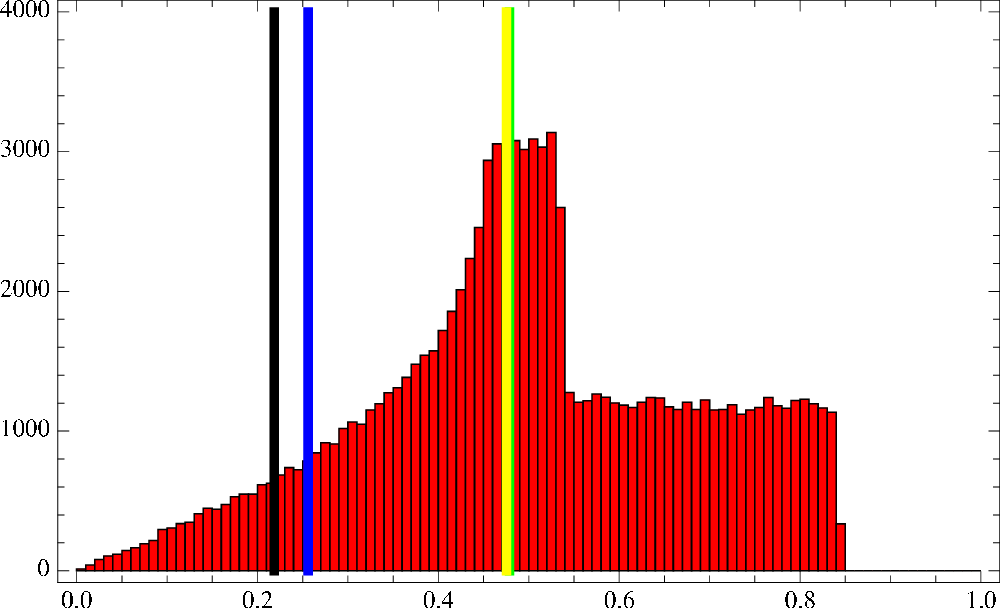}
\caption{$S_{21}$. Vertical lines represent specific directions. We use the following conventions: black for the dipole, green for the north Ecliptic pole, blue for
the north Galactic pole and yellow for the north Super Galactic pole. The plot presents the counts (y-axis) versus the statistic (x-axis).}
\label{S21}
\end{figure}

\begin{figure}
\includegraphics[width=8.0cm]{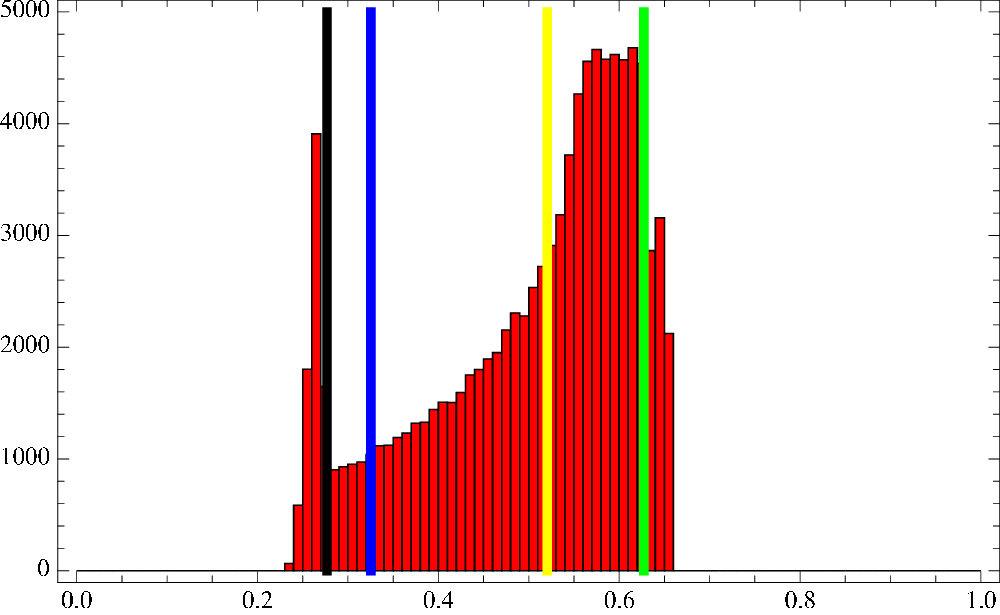}
\caption{$S_{31}$. Same conventions as Fig.~\ref{S21}. The plot presents the counts (y-axis) versus the statistic (x-axis).}
\label{S31}
\end{figure}

\begin{figure}
\includegraphics[width=8.0cm]{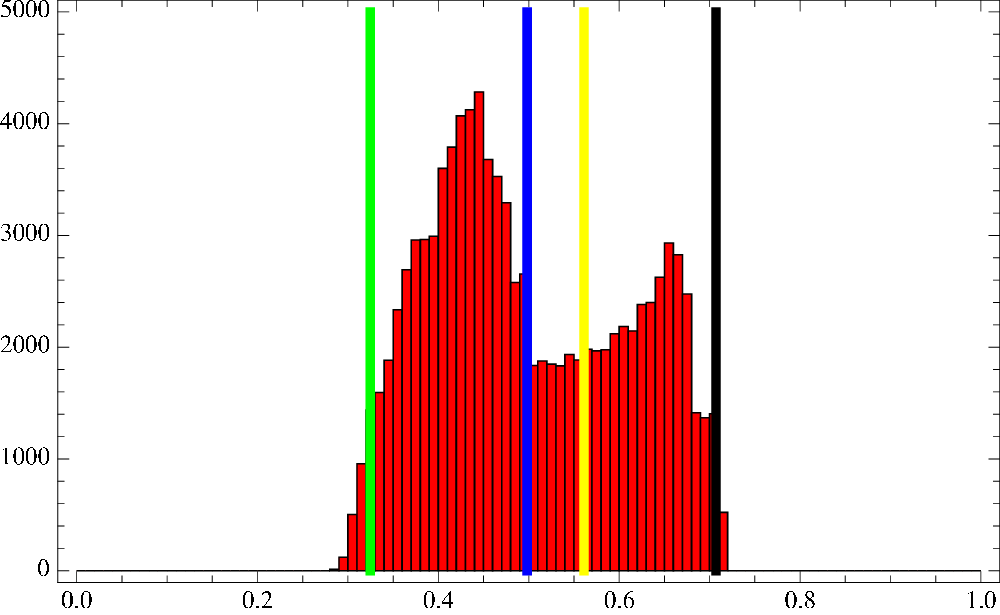}
\caption{$S_{41}$. Same conventions as Fig.~\ref{S21}. The plot presents the counts (y-axis) versus the statistic (x-axis).}
\label{S41}
\end{figure}

\begin{figure}
\includegraphics[width=8.0cm]{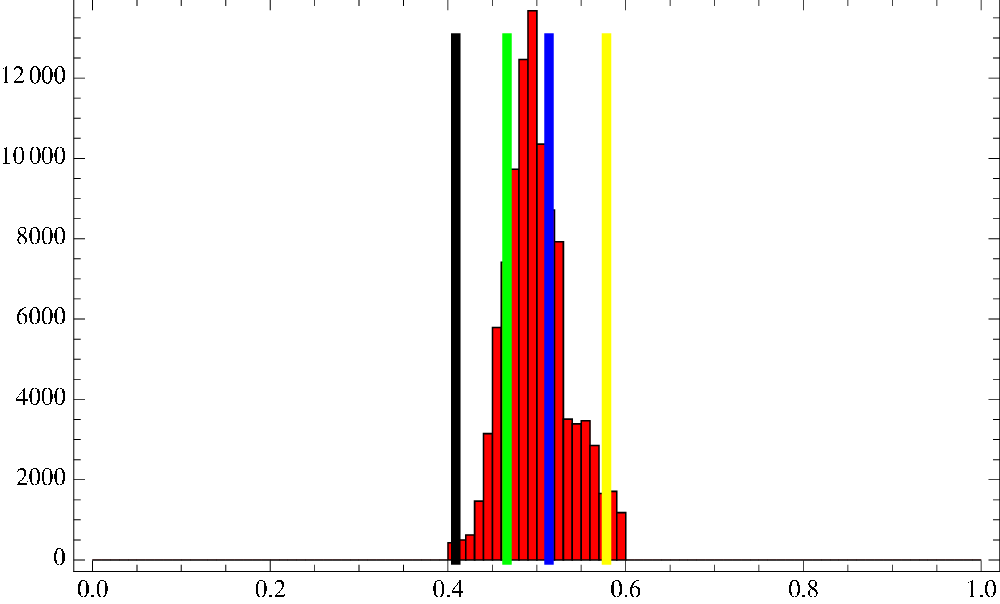}
\caption{$S_{51}$. Same conventions as Fig.~\ref{S21}.. The plot presents the counts (y-axis) versus the statistic (x-axis).}
\label{S51}
\end{figure}

\begin{figure}
\includegraphics[width=8.0cm]{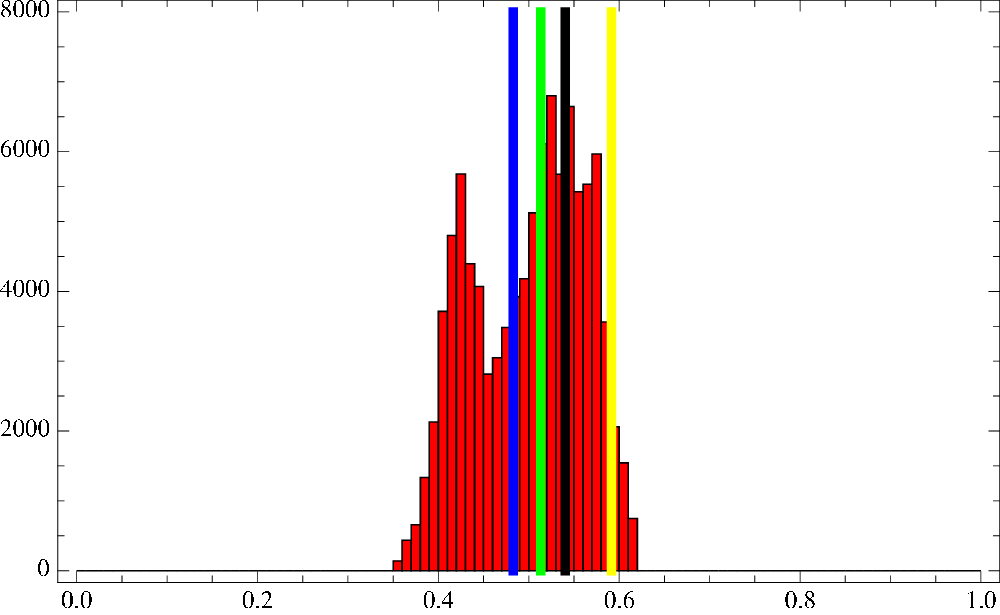}
\caption{$S_{61}$. Same conventions as Fig.~\ref{S21}. The plot presents the counts (y-axis) versus the statistic (x-axis).}
\label{S61}
\end{figure}

\begin{figure}
\includegraphics[width=8.0cm]{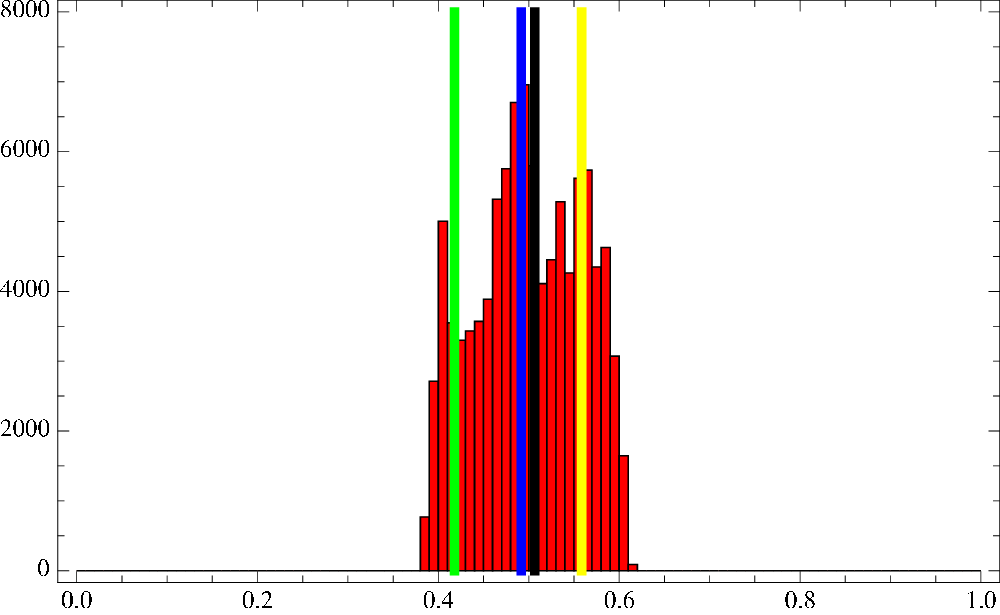}
\caption{$S_{71}$. Same conventions as Fig.~\ref{S21}. The plot presents the counts (y-axis) versus the statistic (x-axis).}
\label{S71}
\end{figure}

\begin{figure}
\includegraphics[width=8.0cm]{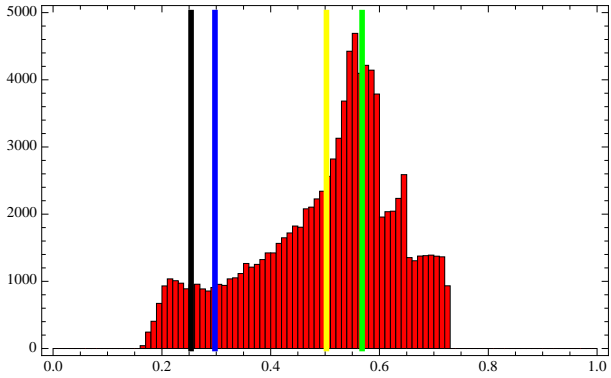}
\caption{$S_{321}^{new}$. Same conventions as Fig.~\ref{S21}. The plot presents the counts (y-axis) versus the statistic (x-axis).}
\label{S321new}
\end{figure}

Results for the second MC are reported in Figures \ref{S231}, \ref{S21}, \ref{S31}, \ref{S41}, \ref{S51}, \ref{S61}, \ref{S71} and \ref{S321new}
where we show the probability distribution function (pdf) for each $S_{\ell 1}$, for $S_{321}$ and for $S_{321}^{new}$,
and the value of the same estimator for the ILC map with various directions, represented by vertical lines in all the plots.
We have chosen the following conventions: black line for the dipole, green line for north Ecliptic pole,
blue line for the north Galactic pole, yellow line for the north Super Galactic pole.
%The values of the estimators of the ILC map with these directions are denoted with DIP, NEP, NGP, NSSP respectively.
In Fig.~\ref{S231} we recover the pdf of $S_{321}$, given in \cite{Copi2006}.
See appendix \ref{shape} for comments about its particular shape.
In Fig.~\ref{S21} we show the pdf for $S_{21}$ and in Fig.~\ref{S31} for $S_{31}$.
Since the DIP-values stand in left tails of the pdfs of both estimators, it is possible to state that the multipole vectors associated to 
quadrupole and octupole are nearly orthogonal to the dipole. The probability to obtain a smaller value is $6.9 \%$ and $7.8 \%$
for $S_{21}$ and $S_{31}$ respectively.
When these estimators are combined as in eq.~(\ref{newS321eq}) than we obtain the pdf shown in Fig.~\ref{S321new}.
In this joint case the probability to obtain a smaller value than the DIP value is $6.6 \%$.
%These (mild) anomalies turn out to be more severe when quadrupole and octupole are combined as in \cite{Copi2006},
%see Fig.~\ref{S231}.
In Fig.~\ref{S41} it is shown the $S_{41}$ estimator.
Since the DIP-value of this estimator is in the right tail of its pdf,
the multipole vectors associated to $\ell=4$ are unlikely aligned
to the dipole. The probability to obtain a larger value is $0.9 \%$. 
In Fig.~\ref{S51} we show the $S_{51}$ estimator.
The DIP-value for this estimator is in the left tail of the corresponding pdf.
Therefore we conclude that the multipole vectors associated to $\ell=5$ are unlikely 
orthogonal to the dipole. The probability to obtain a smaller value is $0.3 \%$. 
In Figs.~\ref{S61} and \ref{S71} we show the pdfs for $S_{61}$ and $S_{71}$
respectively.
In these cases, since the DIP-values fall in the middle of the two pdfs we do not find any anomalies.
For sake of completeness, the probability to obtain a smaller value than the one that is observed is $68.9 \%$ for the $S_{61}$
and $55.1 \%$ for the $S_{71}$.
In Table \ref{tableprobabilities2} the probabilities for each estimator and for each considered direction
are summarized. The most unlikely percentages have been reported in bold.

In Table \ref{tablevectors} we report the multipole vectors associated to the low multipoles of the ILC 5yr map.
For each multipole order $\ell$ we give the $\ell$ multipole vectors writing their colatitude and the longitude (in degrees).

\begin{table}[ht]
\caption{Second MC: probabilities (in percentage) to obtain a smaller value for the considered directions and estimators} % title of Table
\centering % used for centering table
\begin{tabular}{c c c c c} % centered columns (5 columns)
\hline\hline %inserts double horizontal lines
Estimator &  DIP &  NEP &  NGP  &  NSGP \\ [0.5ex] % inserts table
%heading
\hline % inserts single horizontal line
$S_{21}$ & 6.9 & 45.5 & 9.6 & 44.5 \\ % inserting body of the table
$S_{31}$ & 7.8 & 90.8 & 12.4 & 45.7 \\
$S_{321}^{new}$ & 6.6 & 65.7 & 10.5 & 41.7\\
$S_{41}$ & {\bf 99.1} & 2.2 & 55.6 & 67.5 \\
$S_{51}$ & {\bf 0.3} & 16.4 & 69.0 & 96.9 \\
$S_{61}$ & 68.9 & 51.8 & 37.7 & 96.1 \\
$S_{71}$ & 55.1 & 11.3 & 45.3 & 79.4 \\  [1ex] % [1ex] adds vertical space
\hline %inserts single line
\end{tabular}
\label{tableprobabilities2} % is used to refer this table in the text
\end{table}

\begin{table}[ht]
\caption{Multipole vectors for the ILC 5yr map at low $\ell$} % title of Table
\centering % used for centering table
\begin{tabular}{c c c} % centered columns (3 columns)
\hline\hline %inserts double horizontal lines
Multipole order &  Colatitude ($^{\circ}$) &  Longitude  ($^{\circ}$) \\ [0.5ex] % inserts table
%heading
\hline % inserts single horizontal line
$\ell=2$ & 102.03 & 305.16 \\ % inserting body of the table
$\ell=2$ & 106.96 & 184.22  \\
\hline
$\ell=3$ & 77.98 & 313.59 \\
$\ell=3$ & 130.23 & 272.81  \\
$\ell=3$ & 97.18 & 203.91 \\
\hline
$\ell=4$ & 25.05 & 203.82  \\  
$\ell=4$ & 118.63 & 154.69 \\ 
$\ell=4$ & 90.00 & 68.91  \\ 
$\ell=4$ & 128.68 & 37.97  \\ 
\hline
$\ell=5$ & 57.20 & 288.28  \\  
$\ell=5$ & 52.83 & 39.38  \\ 
$\ell=5$ & 125.69 & 279.84 \\ 
$\ell=5$ & 37.20 & 228.60  \\ 
$\ell=5$ & 92.39 & 355.78  \\  
\hline
$\ell=6$ & 37.20 & 34.20  \\ 
$\ell=6$ & 54.32 & 77.34  \\ 
$\ell=6$ & 73.04 & 288.28  \\ 
$\ell=6$ & 81.61 & 222.19  \\  
$\ell=6$ & 32.60 & 239.32  \\ 
$\ell=6$ & 105.71 & 151.88  \\ 
\hline
$\ell=7$ & 55.77 & 345.94  \\ 
$\ell=7$ & 69.26 & 92.81  \\ 
$\ell=7$ & 97.18 & 203.91  \\  
$\ell=7$ & 48.19 & 7.03  \\ 
$\ell=7$ & 62.72 & 174.78  \\ 
$\ell=7$ & 157.94 & 45.00  \\ 
$\ell=7$ & 69.26 & 286.88  \\   [1ex] % [1ex] adds vertical space
\hline %inserts single line
\end{tabular}
\label{tablevectors} % is used to refer this table in the text
\end{table}

%\newpage

\section{Conclusions}
\label{conclusions}

Taking into account a new estimator defined in eq.~(\ref{newS321eq}), we have confirmed
the unlikely alignment of quadrupole and octupole with the dipole in the WMAP 5yr ILC map.
Moreover we have shown that there are new directional anomalies.
%We have shown that there are new directional anomalies beyond the so-called unlikely alignment of quadrupole and octupole 
%with the dipole in the WMAP 5yr ILC map. Such alignment has been confirmed taking into account a new estimator defined in 
%eq.~(\ref{newS321eq}).
Multipole vectors associated to $\ell=4$ are unlikely aligned with the dipole at $99.1 \%$ C.L..
Note that this anomaly holds for both the MCs. 
Furthermore the multipole vectors associated to $\ell=5$ are unlikely orthogonal to the dipole itself,
with a probability at  $99. 7  \%$ C.L. when the MC is performed considering frozen the relative alignments of the low multipoles.
This has been shown through MC simulations that properly take into account WMAP's anisotropic pixel 
noise level (see footnote 5).

What causes these directional anomalies? 
It is difficult to answer this question. 
It is still unknown whether these anomalies come from fundamental physics or whether they 
are the residual of some not perfectly removed astrophysical foreground or systematic effect \cite{abramo2,Naselsky:2006mt,Inoue2006,Inoue2007,cooray2005,Peiris:2010jd,Francis:2009pt}.
As an example of the latter kind, in references \cite{DSCburigana,DSCgruppuso} it is presented a study about the impact
of the dipole straylight contamination
on the low amplitude of the quadrupole and on the low $\ell$ alignments
for {\it Planck}~\footnote{http://www.rssd.esa.int/planck}
characteristics and capabilities.

%Of course such anomalies might be also due to fundamental physics dipolar modulation as studied in \cite{Hanson:2009gu}.
Of course these anomalies might come from fundamental physics, e.g. a non trivial topology of the universe \cite{Bielewicz:2008ga}
or magnetic fields \cite{Bernui:2008ve}. See also \cite{Hanson:2009gu} where a Quadratic Maximum Likelihood method has been adopted to
study a dipolar modulation.

As far as we know few attempts have been tried to study alignments anomalies in polarization \cite{Dvorkin:2007jp} with WMAP data \cite{Frommert:2009qw}. 
New data (i.e. Planck data) are awaited with great interest.
%It will be of great interest to repeat such analysis when {\it Planck} data will be available both in Temperature and Polarization.

%\newpage

\begin{acknowledgments}
We acknowledge the use of the Legacy Archive for Microwave Background Data Analysis (LAMBDA). Support for LAMBDA is provided by the NASA Office of Space Science.
Some of the results in this paper have been derived using the HEALPix \cite{gorski} package.
We acknowledge the use of the public code for the multipole vectors decomposition
(quoted in footnote number $2$) \cite{copi2004}.

This work has been done in the framework of the {\it Planck} LFI activities.
A.~G.  acknowledges support by ASI through ASI/INAF Agreement I/072/09/0 for
the Planck LFI Activity of Phase E2 and I/016/07/0 ÒCOFISÓ.
%A.~G. acknowledges the support by the ASI contracts ÒPlanck LFI ActivityÓ and I/016/07/0 ÒCOFISÓ.

A.~G. warmly thanks P.~Natoli for stimulating and fruitful conversations.
A.~G. wishes to thank JPL people for the kind hospitality during the period in which this work began.
\end{acknowledgments}

\appendix

\section{Some technicalities about MC simulations}
\label{appendix}

In this Appendix we deal with some technicalities about the performed simulations. 
In particular we focus on the second aforementioned MC in which the relative alignments of the low multipoles
are fixed and given by the ILC map. Subsections \ref{shape} and \ref{technicalities} refer to such a MC.

\subsection{Shape of the estimators}
\label{shape}

\begin{figure}
\includegraphics[width=8.0cm]{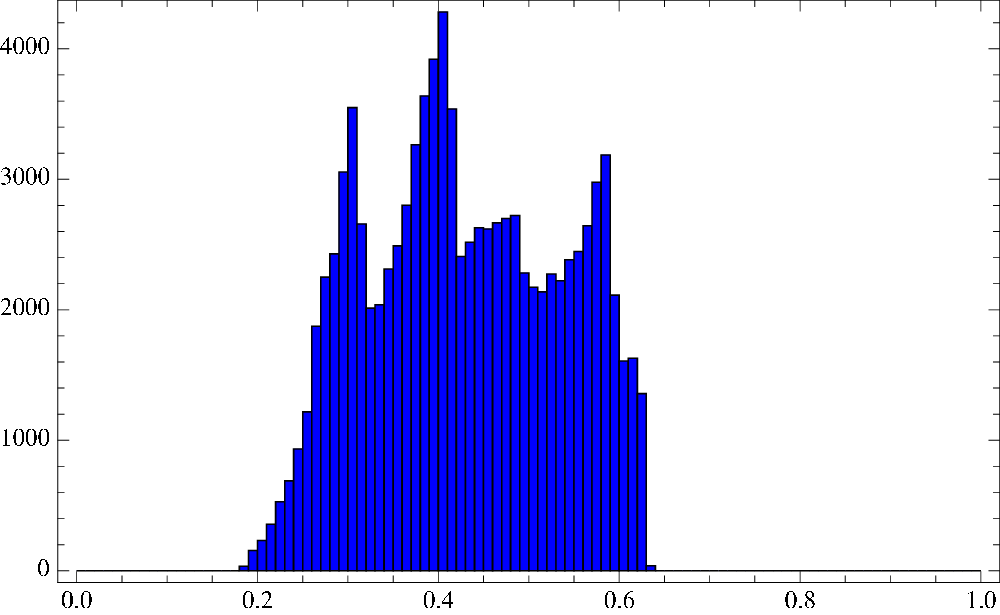}
\caption{Bootstrapping on $S_{321}$ estimator. The plot presents the counts (y-axis) versus the statistic (x-axis).}
\label{S321bs}
\end{figure}

The shape of the pdfs of the estimators is due to the relative alignments of the multipole considered.
In this Section we perform a sort of ``bootstrap'' to the ILC map in order not to change its angular power spectrum
while obtaining a new map with different relative alignments among the multipole vectors.
As an example, we recompute the pdf of $S_{321}$ for this new map. This is done just to show
that the particular (and sometimes puzzling) shape of the pdf of the considered estimators strongly depend on the phases
of map taken under analysis. In order to not to change the angular power spectrum, we arbitrarily reshuffle the $a_{\ell m}$ for $\ell=2,3$ of the ILC map as follows:
$$a_{2 \,\,1/2}^{ILC} \rightarrow a_{2 \,\, 2/1}^{ILC}$$ and $$a_{3 \,\,1/2/3}^{ILC} \rightarrow a_{2 \,\, 2/3/1}^{ILC}$$
leaving $a_{2,3 \,\,0}^{ILC}$ unchanged.
In Fig.~\ref{S321bs} we show how the shape of the pdf for the estimator $S_{321}$ change.

\subsection{Impact of the degradation procedure}
\label{technicalities}

\begin{figure}
\includegraphics[width=8.0cm]{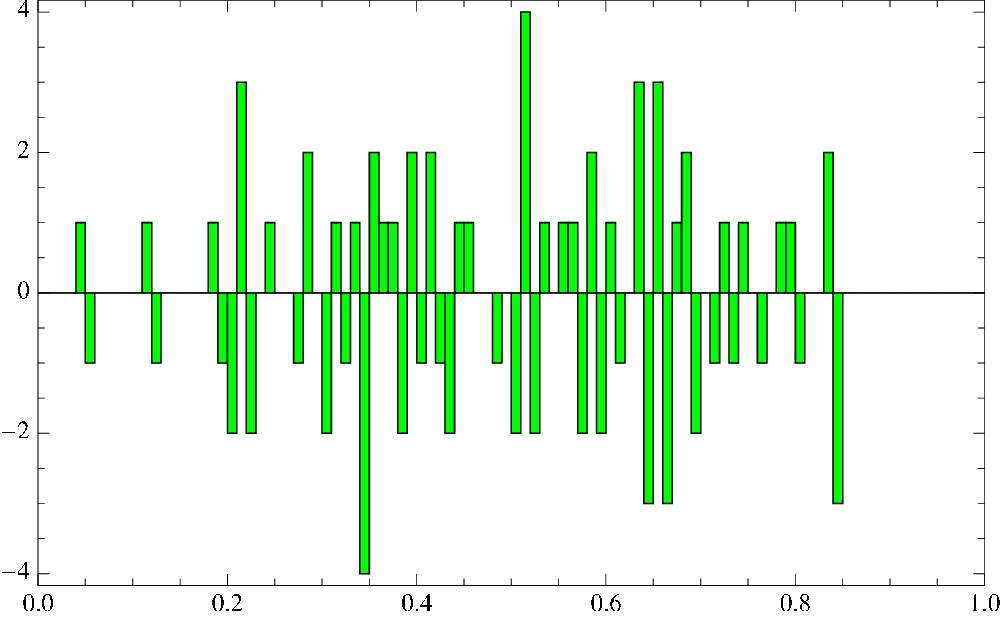} 
\includegraphics[width=8.0cm]{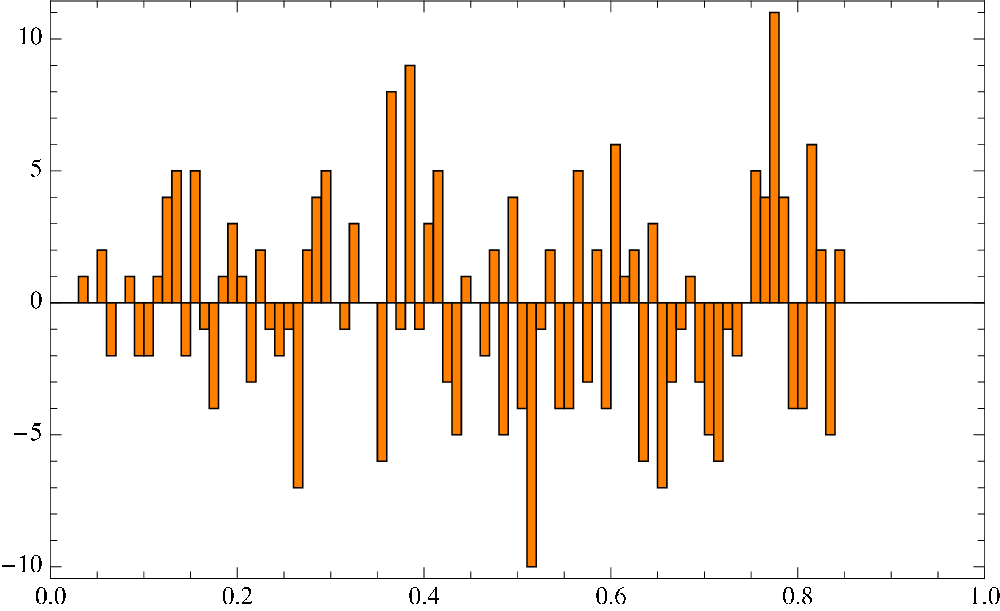}
\caption{Left Panel: Degradation effect on $S_{21}$. Right Panel: Noise effect on $S_{21}$.
Both the panels present the counts (y-axis) versus the statistic (x-axis).}
%\label{diff_hr_lr_S21}
%\end{figure}
%\begin{figure}
%\includegraphics[width=8.0cm]{S21_noise_effect.pdf}
%\caption{Noise effect on $S_{21}$. The plot presents the counts (y-axis) versus the statistic (x-axis).}
%\label{diff_lrsn_lrs_S21}
\label{S21test}
\end{figure}

In this Section we briefly discuss our MC inputs.
The ILC map is provided by the WMAP team at $N_{side}=512$
%\footnote{The Healpix parameter $N_{side}$ is related to the number of pixels $N_{pix}$ through $N_{pix}=12 \, N_{side}^2$}
and has to be degraded to be handled in a MC. 
We chose to work at $N_{side}=16$ and we have to 
keep under control the impact of this degradation\footnote{Since we are interested in quantities that depend on phases rather than norms
of $a_{\ell m}$, the impact of the degradation upon angular power spectrum is not
what we have to study. However, for sake of completeness,  
the effect of this degradation on angular power spectrum is of the order of
few $\mu K^2$ that is well below the cosmic variance but above the
angular power spectrum of the WMAP noise expected at those angular scales.}.
In order to evaluate the effect of the degradation on our results, we have
compared the pdf's of the considered estimators obtained performing a MC directly at high resolution 
(i.e. $N_{side}=512$) to the ones obtained by a (far less expensive) MC at low resolution (i.e. $N_{side}=16$).
Since the high resolution MC is computationally heavy, we have 
reduced the number of random (angles) extractions to $10^3$.
As an example, in the left panel of Fig.~\ref{S21test} we show the difference between the histogram at high resolution
and the histogram at low resolution for the estimator $S_{21}$. 
This plot of the difference has to be compared
to the difference between the histogram at low resolution that include an
(anisotropic) noise realization (at the V band WMAP level) and
the histogram at low resolution (just degraded, without any noise realization on it).
See right panel of Fig.~\ref{S21test} for the noise effect on the estimator $S_{21}$.
The level of the fluctuations in the right panel of Fig.~\ref{S21test}
is larger than in the left panel of the same Figure,
%.~\ref{diff_hr_lr_S21}
showing that the effect of noise
covers the effect of the degradation. 
In order to quantify degradation
and noise effects, we compute the standard deviation 
of the two histograms of the differences,
obtaining
$0.353$ for the noise impact, and $0.136$ for the degradation impact.
See Table \ref{degradation_vs_noise} for a summary of the above defined 
standard deviation for all the considered estimators.
Since the effect of V band WMAP (anisotropic) noise is larger than the
effect of the degradation for all the considered cases, we conclude that the process of the degradation
does not introduce any significant spurious effect on the estimators we used to study anomalous directions.

\begin{table}[ht]
\caption{Impact of noise vs degradation} % title of Table
\centering % used for centering table
\begin{tabular}{c c c} % centered columns (2 columns)
\hline\hline %inserts double horizontal lines
Estimator & $\sigma_{Noise}$  &  $\sigma_{Degrad}$  \\ [0.5ex] % inserts table
%heading
\hline % inserts single horizontal line
$S_{21}$ & 0.353 & 0.136  \\ % inserting body of the table
$S_{31}$ & 0.444 & 0.066  \\
$S_{41}$ & 0.403 & 0.108 \\
$S_{51}$ & 0.242 & 0.107 \\
$S_{61}$ & 0.573 & 0.191  \\
$S_{71}$ & 0.512 & 0.132  \\  [1ex] % [1ex] adds vertical space
\hline %inserts single line
\end{tabular}
\label{degradation_vs_noise} % is used to refer this table in the text
\end{table}

%\begin{widetext}

%\end{widetext}


\begin{thebibliography}{99}
\bibitem{WMAP5yr}
  E.~Komatsu {\it et al.}  [WMAP Collaboration],
  %``Five-Year Wilkinson Microwave Anisotropy Probe (WMAP\altaffilmark 1 )
  %Observations:Cosmological Interpretation,''
  arXiv:0803.0547 [astro-ph].
  %%CITATION = ARXIV:0803.0547;%%
  J.~Dunkley {\it et al.}  [WMAP Collaboration],
  %``Five-Year Wilkinson Microwave Anisotropy Probe (WMAP) Observations:
  %Likelihoods and Parameters from the WMAP data,''
  arXiv:0803.0586 [astro-ph].
  %%CITATION = ARXIV:0803.0586;%%

\bibitem{copi2004}
C.~J.~Copi, D.~Huterer and G.~D.~Starkman,
  %``Multipole Vectors--a new representation of the CMB sky and evidence for
  %statistical anisotropy or non-Gaussianity at 2<=l<=8,''
  Phys.\ Rev.\  D {\bf 70}, 043515 (2004)
  [arXiv:astro-ph/0310511].
  %%CITATION = PHRVA,D70,043515;%%
 
 \bibitem{weeks}
  J.~R.~Weeks,
  %``Maxwell's Multipole Vectors and the CMB,''
  arXiv:astro-ph/0412231.
  %%CITATION = ASTRO-PH/0412231;%%

\bibitem{lowalignments}
  M.~Tegmark, A.~de Oliveira-Costa and A.~Hamilton,
  %``A high resolution foreground cleaned CMB map from WMAP,''
  Phys.\ Rev.\  D {\bf 68}, 123523 (2003)
  [arXiv:astro-ph/0302496].
  %%CITATION = PHRVA,D68,123523;%%
%% C.~J.~Copi, D.~Huterer and G.~D.~Starkman,
 %% %``Multipole Vectors--a new representation of the CMB sky and evidence for
 %% %statistical anisotropy or non-Gaussianity at 2<=l<=8,''
 %% Phys.\ Rev.\  D {\bf 70}, 043515 (2004)
 %% [arXiv:astro-ph/0310511].
 %% %%CITATION = PHRVA,D70,043515;%%
D.~J.~Schwarz, G.~D.~Starkman, D.~Huterer and C.~J.~Copi,
  %``Is the low-l microwave background cosmic?,''
  Phys.\ Rev.\ Lett.\  {\bf 93}, 221301 (2004)
  [arXiv:astro-ph/0403353].
  %%CITATION = PRLTA,93,221301;%%
  K.~Land and J.~Magueijo,
  %``The axis of evil,''
  Phys.\ Rev.\ Lett.\  {\bf 95}, 071301 (2005)
  [arXiv:astro-ph/0502237].
  %%CITATION = PRLTA,95,071301;%%
C.~Vale,
  %``Local Pancake Defeats Axis of Evil,''
  arXiv:astro-ph/0509039.
  %%CITATION = ASTRO-PH/0509039;%%
 %% J.~R.~Weeks,
 %% %``Maxwell's Multipole Vectors and the CMB,''
 %% arXiv:astro-ph/0412231.
 %% %%CITATION = ASTRO-PH/0412231;%%


%\bibitem{Copi2008}
%C.~J.~Copi, D.~Huterer, D.~J.~Schwarz and G.~D.~Starkman,
%  %``No large-angle correlations on the non-Galactic microwave sky,''
%  arXiv:0808.3767 [astro-ph].
%  %%CITATION = ARXIV:0808.3767;%%

\bibitem{CopiMNRAS2006}
  C.~J.~Copi, D.~Huterer, D.~J.~Schwarz and G.~D.~Starkman,
  %``On the large-angle anomalies of the microwave sky,''
  Mon.\ Not.\ Roy.\ Astron.\ Soc.\  {\bf 367}, 79 (2006)
  [arXiv:astro-ph/0508047].
  %%CITATION = MNRAA,367,79;%%

%\cite{Gruppuso:2009ee}
\bibitem{Gruppuso:2009ee}
  A.~Gruppuso and C.~Burigana,
  %``Large scale alignment anomalies of CMB anisotropies: a new test for
  %residuals applied to WMAP 5yr maps,''
  JCAP {\bf 0908}, 004 (2009)
  [arXiv:0907.1949 [astro-ph.CO]].
  %%CITATION = JCAPA,0908,004;%%

\bibitem{Copi2006}
C.~Copi, D.~Huterer, D.~Schwarz and G.~Starkman,
  %``The Uncorrelated Universe: Statistical Anisotropy and the Vanishing Angular
  %Correlation Function in WMAP Years 1-3,''
  Phys.\ Rev.\  D {\bf 75}, 023507 (2007)
  [arXiv:astro-ph/0605135].
  %%CITATION = PHRVA,D75,023507;%%

%\cite{Copi:2005ff}
%\bibitem{CopiMNRAS2006}
%  C.~J.~Copi, D.~Huterer, D.~J.~Schwarz and G.~D.~Starkman,
%  %``On the large-angle anomalies of the microwave sky,''
%  Mon.\ Not.\ Roy.\ Astron.\ Soc.\  {\bf 367}, 79 (2006)
%  [arXiv:astro-ph/0508047].
%  %%CITATION = MNRAA,367,79;%%

%\bibitem{Smoot1992}
%G.~F.~Smoot {\it et al.},
%  %``Structure in the COBE differential microwave radiometer first year maps,''
%  Astrophys.\ J.\  {\bf 396}, L1 (1992).
%  %%CITATION = ASJOA,396,L1;%%

%\bibitem{Hinshaw1996}
%  G.~Hinshaw, A.~J.~Banday, C.~L.~Bennett, K.~M.~Gorski, A.~Kogut, G.~F.~Smoot and E.~L.~Wright,
%  %``Band Power Spectra in the COBE DMR 4-Year Anisotropy Maps,''
%  arXiv:astro-ph/9601058.
%  %%CITATION = ASTRO-PH/9601058;%%

%\bibitem{copi2004}
%C.~J.~Copi, D.~Huterer and G.~D.~Starkman,
%  %``Multipole Vectors--a new representation of the CMB sky and evidence for
%  %statistical anisotropy or non-Gaussianity at 2<=l<=8,''
%  Phys.\ Rev.\  D {\bf 70}, 043515 (2004)
%  [arXiv:astro-ph/0310511].
%  %%CITATION = PHRVA,D70,043515;%%
% 
% \bibitem{weeks}
%  J.~R.~Weeks,
%  %``Maxwell's Multipole Vectors and the CMB,''
%  arXiv:astro-ph/0412231.
%  %%CITATION = ASTRO-PH/0412231;%%

%\bibitem{lowalignments}
%  M.~Tegmark, A.~de Oliveira-Costa and A.~Hamilton,
%  %``A high resolution foreground cleaned CMB map from WMAP,''
%  Phys.\ Rev.\  D {\bf 68}, 123523 (2003)
%  [arXiv:astro-ph/0302496].
%  %%CITATION = PHRVA,D68,123523;%%
%%% C.~J.~Copi, D.~Huterer and G.~D.~Starkman,
% %% %``Multipole Vectors--a new representation of the CMB sky and evidence for
% %% %statistical anisotropy or non-Gaussianity at 2<=l<=8,''
% %% Phys.\ Rev.\  D {\bf 70}, 043515 (2004)
% %% [arXiv:astro-ph/0310511].
% %% %%CITATION = PHRVA,D70,043515;%%
%D.~J.~Schwarz, G.~D.~Starkman, D.~Huterer and C.~J.~Copi,
%  %``Is the low-l microwave background cosmic?,''
%  Phys.\ Rev.\ Lett.\  {\bf 93}, 221301 (2004)
%  [arXiv:astro-ph/0403353].
%  %%CITATION = PRLTA,93,221301;%%
%  K.~Land and J.~Magueijo,
%  %``The axis of evil,''
%  Phys.\ Rev.\ Lett.\  {\bf 95}, 071301 (2005)
%  [arXiv:astro-ph/0502237].
%  %%CITATION = PRLTA,95,071301;%%
%C.~Vale,
%  %``Local Pancake Defeats Axis of Evil,''
%  arXiv:astro-ph/0509039.
%  %%CITATION = ASTRO-PH/0509039;%%
% %% J.~R.~Weeks,
% %% %``Maxwell's Multipole Vectors and the CMB,''
% %% arXiv:astro-ph/0412231.
% %% %%CITATION = ASTRO-PH/0412231;%%

\bibitem{abramo}
L.~R.~Abramo, A.~Bernui, I.~S.~Ferreira, T.~Villela and C.~A.~Wuensche,
  %``Alignment Tests for low CMB multipoles,''
  Phys.\ Rev.\  D {\bf 74}, 063506 (2006)
  [arXiv:astro-ph/0604346].
  %%CITATION = PHRVA,D74,063506;%%

%\bibitem{Eriksen}
%H.~K.~Eriksen, F.~K.~Hansen, A.~J.~Banday, K.~M.~Gorski and P.~B.~Lilje,
%  %``Asymmetries in the CMB anisotropy field,''
%  Astrophys.\ J.\  {\bf 605}, 14 (2004)
%  [Erratum-ibid.\  {\bf 609}, 1198 (2004)]
%  [arXiv:astro-ph/0307507].
%  %%CITATION = ASJOA,605,14;%%
  
%\bibitem{ForegroundDataAnalisys}

\bibitem{Katz}
  G.~Katz and J.~Weeks,
  %``Polynomial Interpretation of Multipole Vectors,''
  Phys.\ Rev.\  D {\bf 70}, 063527 (2004)
  [arXiv:astro-ph/0405631].
  %%CITATION = PHRVA,D70,063527;%%


%\bibitem{methods}
%%\bibitem{Coles:2003dw}
%  P.~Coles, P.~Dineen, J.~Earl and D.~Wright,
%  %``Phase Correlations in Cosmic Microwave Background Temperature Maps,''
%  Mon.\ Not.\ Roy.\ Astron.\ Soc.\  {\bf 350}, 983 (2004)
%  [arXiv:astro-ph/0310252].
%  %%CITATION = MNRAA,350,983;%%
%%bibitem{Park:2006dv}
%  C.~G.~Park, C.~Park and J.~R.~I.~Gott,
%  %``Cleaned Three-Year WMAP CMB Map: Magnitude of the Quadrupole and
%  %Alignment of Large Scale Modes,''
%  Astrophys.\ J.\  {\bf 660}, 959 (2007)
%  [arXiv:astro-ph/0608129].
%  %%CITATION = ASJOA,660,959;%%
%%\bibitem{Naselsky:2007gt}
%  P.~D.~Naselsky, O.~V.~Verkhodanov and M.~T.~B.~Nielsen,
%  %``Instability of reconstruction of the low CMB multipoles,''
%  Astrophys.\ Bull.\  {\bf 63}, 216 (2008)
%  [arXiv:0707.1484 [astro-ph]].
%  %%CITATION = 00588,63,216;%%
%%\bibitem{Chiang:2007rp}
%  L.~Y.~Chiang, P.~D.~Naselsky and P.~Coles,
%  %``Cosmic Covariance and the Low Quadrupole Anisotropy of the Wilkinson
%  %Microwave Anisotropy Probe (WMAP) Data,''
%  Mod.\ Phys.\ Lett.\  A {\bf 23}, 1489 (2008)
%  [arXiv:0711.1860 [astro-ph]].
%  %%CITATION = MPLAE,A23,1489;%%

%\bibitem{FundamentalPhysics}
%%\bibitem{Luminet:2003dx}
%  J.~P.~Luminet, J.~Weeks, A.~Riazuelo, R.~Lehoucq and J.~P.~Uzan,
%  %``Dodecahedral space topology as an explanation for weak wide-angle
%  %temperature correlations in the cosmic microwave background,''
%  Nature {\bf 425}, 593 (2003)
%  [arXiv:astro-ph/0310253].
%  %%CITATION = NATUA,425,593;%%
%%bibitem{Weeks:2006rr}
%  J.~R.~Weeks and J.~Gundermann,
%  %``Dodecahedral topology fails to explain quadrupole-octupole alignment,''
%  Class.\ Quant.\ Grav.\  {\bf 24}, 1863 (2007)
%  [arXiv:astro-ph/0611640].
%  %%CITATION = CQGRD,24,1863;%%
%  %\bibitem{Contaldi:2003zv}
%  C.~R.~Contaldi, M.~Peloso, L.~Kofman and A.~Linde,
%  %``Suppressing the lower Multipoles in the CMB Anisotropies,''
%  JCAP {\bf 0307} (2003) 002
%  [arXiv:astro-ph/0303636].
%  %%CITATION = JCAPA,0307,002;%%
%%\bibitem{Moroi:2003pq}
%  T.~Moroi and T.~Takahashi,
%  %``Correlated isocurvature fluctuation in quintessence and suppressed CMB
%  %anisotropies at low multipoles,''
%  Phys.\ Rev.\ Lett.\  {\bf 92} (2004) 091301
%  [arXiv:astro-ph/0308208].
%  %%CITATION = PRLTA,92,091301;%%
%L.~Campanelli, P.~Cea and L.~Tedesco,
%  %``Ellipsoidal Universe Can Solve The CMB Quadrupole Problem,''
%  Phys.\ Rev.\ Lett.\  {\bf 97}, 131302 (2006)
%  [Erratum-ibid.\  {\bf 97}, 209903 (2006)]
%  [arXiv:astro-ph/0606266].
%  %%CITATION = PRLTA,97,131302;%%
%A.~Gruppuso,
%  %``A Complete Statistical Analysis for the Quadrupole Amplitude in an
%  %Ellipsoidal Universe,''
%  Phys.\ Rev.\  D {\bf 76}, 083010 (2007)
%  [arXiv:0705.2536 [astro-ph]].
%  %%CITATION = PHRVA,D76,083010;%%
%  %\cite{Finelli:2005zc}
%%\bibitem{Finelli:2005zc}
%  F.~Finelli and A.~Gruppuso,
%  %``A remark on 'A CMB / dark energy cosmic duality',''
%  arXiv:hep-th/0501089.
%  %%CITATION = HEP-TH/0501089;%%
%  %J.~D.~Barrow, P.~G.~Ferreira and J.~Silk,
%  %``Constraints on a Primordial Magnetic Field,''
%  %Phys.\ Rev.\ Lett.\  {\bf 78}, 3610 (1997)
%  %[arXiv:astro-ph/9701063]. [[?]]
%  %%CITATION = PRLTA,78,3610;%%
%  %T.~Ghosh, A.~Hajian and T.~Souradeep,
%  %``Unveiling Hidden Patterns in CMB Anisotropy Maps,''
%  %Phys.\ Rev.\  D {\bf 75}, 083007 (2007)
%  %[arXiv:astro-ph/0604279].
%  %%CITATION = PHRVA,D75,083007;%%
%  %T.~R.~Jaffe, A.~J.~Banday, H.~K.~Eriksen, K.~M.~Gorski and F.~K.~Hansen,
%  %``Bianchi Type VII_h Models and the WMAP 3-year Data,''
%  %Astron.\ Astrophys.\  {\bf 460}, 393 (2006)
%  %[arXiv:astro-ph/0606046].
%  %%CITATION = AAEJA,460,393;%%
%  % \bibitem{de OliveiraCosta:2003pu}
%  A.~de Oliveira-Costa, M.~Tegmark, M.~Zaldarriaga and A.~Hamilton,
%  %``The significance of the largest scale CMB fluctuations in WMAP,''
%  Phys.\ Rev.\  D {\bf 69}, 063516 (2004)
%  [arXiv:astro-ph/0307282].
%  %%CITATION = PHRVA,D69,063516;%%
  
%\bibitem{Bunn}
%  E.~F.~Bunn and A.~Bourdon,
%  %``Contamination cannot explain the lack of large-scale power in the cosmic
%  %microwave background radiation,''
%  arXiv:0808.0341 [astro-ph].
%  %%CITATION = ARXIV:0808.0341;%%
%  
%\bibitem{bielewicz_2}
%P.~Bielewicz, H.~K.~Eriksen, A.~J.~Banday, K.~M.~Gorski and P.~B.~Lilje,
%  %``Multipole vector anomalies in the first-year WMAP data: a cut-sky
%  %analysis,''
%  Astrophys.\ J.\  {\bf 635}, 750 (2005)
%  [arXiv:astro-ph/0507186].
%  %%CITATION = ASJOA,635,750;%%

%\bibitem{bielewicz_1}
% P.~Bielewicz, K.~M.~Gorski and A.~J.~Banday,
%  %``Low order multipole maps of CMB anisotropy derived from WMAP,''
%  Mon.\ Not.\ Roy.\ Astron.\ Soc.\  {\bf 355}, 1283 (2004)
%  [arXiv:astro-ph/0405007].
%  %%CITATION = MNRAA,355,1283;%%

%\bibitem{deOliveiraCosta:2006}
%  A.~de Oliveira-Costa and M.~Tegmark,
%  %``CMB multipole measurements in the presence of foregrounds,''
%  Phys.\ Rev.\  D {\bf 74}, 023005 (2006)
%  [arXiv:astro-ph/0603369].
%  %%CITATION = PHRVA,D74,023005;%%

%\bibitem{Katz}
%  G.~Katz and J.~Weeks,
%  %``Polynomial Interpretation of Multipole Vectors,''
%  Phys.\ Rev.\  D {\bf 70}, 063527 (2004)
%  [arXiv:astro-ph/0405631].
%  %%CITATION = PHRVA,D70,063527;%%

\bibitem{gorski}
K.M. Gorski, E. Hivon, A.J. Banday, B.D. Wandelt, F.K. Hansen, M. Reinecke, and M. Bartelmann,
HEALPix: A Framework for High-resolution Discretization and Fast Analysis of Data Distributed on the Sphere, Ap.J., 622, 759-771, 2005.

%\bibitem{deOliveiraCosta:2006}
 % A.~de Oliveira-Costa and M.~Tegmark,
  %%``CMB multipole measurements in the presence of foregrounds,''
  %Phys.\ Rev.\  D {\bf 74}, 023005 (2006)
  %[arXiv:astro-ph/0603369].
  %%CITATION = PHRVA,D74,023005;%%

\bibitem{landmagueijo}
  K.~Land and J.~Magueijo,
  %``The Multipole Vectors of WMAP, and their frames and invariants,''
  Mon.\ Not.\ Roy.\ Astron.\ Soc.\  {\bf 362}, 838 (2005)
  [arXiv:astro-ph/0502574].
  %%CITATION = MNRAA,362,838;%%

\bibitem{dennis}
M.~R.~Dennis, J. Phys. A 37 (2004) 9487-9500, 
J.Phys. A38 (2005) 1653-1658

\bibitem{abramo2}
L.~R.~Abramo, L.~S.~Jr. and C.~A.~Wuensche,
 %``Anomalies in the low CMB multipoles and extended foregrounds,''
 Phys.\ Rev.\  D {\bf 74}, 083515 (2006)
   [arXiv:astro-ph/0605269].
    %%CITATION = PHRVA,D74,083515;%%

\bibitem{Naselsky:2006mt}
P.~D.~Naselsky and O.~V.~Verkhodanov,
 %``Peculiarities of phases of the WMAP quadrupole,''
Int.\ J.\ Mod.\ Phys.\  D {\bf 17}, 179 (2008)
 [arXiv:astro-ph/0609409].
 %%CITATION = IMPAE,D17,179;%%

\bibitem{Inoue2006}
K.~T.~Inoue and J.~Silk,
  %``Local Voids as the Origin of Large-angle Cosmic Microwave Background
  %Anomalies,''
  Astrophys.\ J.\  {\bf 648}, 23 (2006)
  [arXiv:astro-ph/0602478].
  %%CITATION = ASJOA,648,23;%%

\bibitem{Inoue2007}
   K.~T.~Inoue and J.~Silk,
  %``Local Voids as the Origin of Large-angle Cosmic Microwave Background
  %Anomalies: The Effect of a Cosmological Constant,''
  Astrophys.\ J.\  {\bf 664}, 650 (2007)
  [arXiv:astro-ph/0612347].
  %%CITATION = ASJOA,664,650;%%

\bibitem{cooray2005}    
A.~Cooray and N.~Seto,
  %``Did WMAP see Moving Local Structures?,''
  JCAP {\bf 0512}, 004 (2005)
  [arXiv:astro-ph/0510137].
  %%CITATION = JCAPA,0512,004;%%

\bibitem{Peiris:2010jd}
  H.~V.~Peiris and T.~L.~Smith,
  %``CMB Isotropy Anomalies and the Local Kinetic Sunyaev-Zel'dovich Effect,''
  arXiv:1002.0836 [astro-ph.CO].
  %%CITATION = ARXIV:1002.0836;%%
  
 \bibitem{Francis:2009pt}
  C.~L.~Francis and J.~A.~Peacock,
  %``An estimate of the local ISW signal, and its impact on CMB anomalies,''
  arXiv:0909.2495 [astro-ph.CO].
  %%CITATION = ARXIV:0909.2495;%%

\bibitem{DSCburigana}
C.~Burigana, A.~Gruppuso and F.~Finelli,
  %``On the dipole straylight contamination in spinning space missions dedicated
  %to CMB anisotropy,''
  Mon.\ Not.\ Roy.\ Astron.\ Soc.\  {\bf 371}, 1570 (2006)
  [arXiv:astro-ph/0607506].
  %%CITATION = MNRAA,371,1570;%%

\bibitem{DSCgruppuso}
  A.~Gruppuso, C.~Burigana and F.~Finelli,
  %``The impact of dipole straylight contamination on the alignment of low
  %multipoles of CMB anisotropies,''
  Mon.\ Not.\ Roy.\ Astron.\ Soc.\  {\bf 376} (2007) 907
  [arXiv:astro-ph/0701295].
  %%CITATION = MNRAA,376,907;%%
  
  \bibitem{Hanson:2009gu}
  D.~Hanson and A.~Lewis,
  %``Estimators for CMB Statistical Anisotropy,''
  Phys.\ Rev.\  D {\bf 80}, 063004 (2009)
  [arXiv:0908.0963 [astro-ph.CO]].
  %%CITATION = PHRVA,D80,063004;%%
  
\bibitem{Bielewicz:2008ga}
  P.~Bielewicz and A.~Riazuelo,
  %``The study of topology of the universe using multipole vectors,''
  arXiv:0804.2437 [astro-ph].
  %%CITATION = ARXIV:0804.2437;%%

%\cite{Bernui:2008ve}
\bibitem{Bernui:2008ve}
  A.~Bernui and W.~S.~Hipolito-Ricaldi,
  %``Can a primordial magnetic field originate large-scale anomalies in WMAP
  %data?,''
  Mon.\ Not.\ Roy.\ Astron.\ Soc.\  {\bf 389}, 1453 (2008)
  [arXiv:0807.1076 [astro-ph]].
  %%CITATION = MNRAA,389,1453;%%

 \bibitem{Dvorkin:2007jp}
  C.~Dvorkin, H.~V.~Peiris and W.~Hu,
  %``Testable polarization predictions for models of CMB isotropy anomalies,''
  Phys.\ Rev.\  D {\bf 77} (2008) 063008
  [arXiv:0711.2321 [astro-ph]].
  %%CITATION = PHRVA,D77,063008;%%

\bibitem{Frommert:2009qw}
  M.~Frommert and T.~A.~Ensslin,
  %``The axis of evil - a polarization perspective,''
  arXiv:0908.0453 [astro-ph.CO].
  %%CITATION = ARXIV:0908.0453;%%
  
%
%\bibitem{Hinshaw}
%  G.~Hinshaw {\it et al.}  [WMAP Collaboration],
%  %``Three-year Wilkinson Microwave Anisotropy Probe (WMAP) observations:
%  %Temperature analysis,''
%  arXiv:astro-ph/0603451.
%  %%CITATION = ASTRO-PH/0603451;%%

%\bibitem{Copi2006}
%  C.~Copi, D.~Huterer, D.~Schwarz and G.~Starkman,
%  %``The Uncorrelated Universe: Statistical Anisotropy and the Vanishing Angular
%  %Correlation Function in WMAP Years 1-3,''
%  Phys.\ Rev.\  D {\bf 75}, 023507 (2007)
%  [arXiv:astro-ph/0605135].
%  %%CITATION = PHRVA,D75,023507;%%

%\bibitem{deOliveira-Costa2003}
%  A.~de Oliveira-Costa, M.~Tegmark, M.~Zaldarriaga and A.~Hamilton,
%  %``The significance of the largest scale CMB fluctuations in WMAP,''
%  Phys.\ Rev.\  D {\bf 69}, 063516 (2004)
%  [arXiv:astro-ph/0307282].
%  %%CITATION = PHRVA,D69,063516;%%

%\bibitem{Schwarz2004}
%  D.~J.~Schwarz, G.~D.~Starkman, D.~Huterer and C.~J.~Copi,
%  %``Is the low-l microwave background cosmic?,''
%  Phys.\ Rev.\ Lett.\  {\bf 93}, 221301 (2004)
%  [arXiv:astro-ph/0403353].
%  %%CITATION = PRLTA,93,221301;%%

%\bibitem{Copi2003}
%  C.~J.~Copi, D.~Huterer and G.~D.~Starkman,
%  %``Multipole Vectors--a new representation of the CMB sky and evidence for
%  %statistical anisotropy or non-Gaussianity at 2<=l<=8,''
%  Phys.\ Rev.\  D {\bf 70}, 043515 (2004)
%  [arXiv:astro-ph/0310511].
%  %%CITATION = PHRVA,D70,043515;%%

%\bibitem{Abramo2006}
%  L.~R.~Abramo, A.~Bernui, I.~S.~Ferreira, T.~Villela and C.~A.~Wuensche,
%  %``Alignment Tests for low CMB multipoles,''
%  Phys.\ Rev.\  D {\bf 74}, 063506 (2006)
%  [arXiv:astro-ph/0604346].
%  %%CITATION = PHRVA,D74,063506;%%

%%\cite{Abramo:2006hs}
%\bibitem{Abramo2006bis}
%  L.~R.~Abramo, L.~S.~Jr. and C.~A.~Wuensche,
%  %``Anomalies in the low CMB multipoles and extended foregrounds,''
%  Phys.\ Rev.\  D {\bf 74}, 083515 (2006)
%  [arXiv:astro-ph/0605269].
%  %%CITATION = PHRVA,D74,083515;%%

%%\bibitem{Abramo2003}
%%  L.~R.~Abramo and L.~J.~Sodre,
%  %``Can the Local Supercluster explain de low CMB multipoles?,''
%%  arXiv:astro-ph/0312124.
%  %%CITATION = ASTRO-PH/0312124;%%

%\bibitem{DSCamplitude}

%%\bibitem{Burigana2006}
%  C.~Burigana, A.~Gruppuso and F.~Finelli,
%  %``On the dipole straylight contamination in spinning space missions dedicated
%  %to CMB anisotropy,''
%  Mon.\ Not.\ Roy.\ Astron.\ Soc.\  {\bf 371}, 1570 (2006)
%  [arXiv:astro-ph/0607506].
%  %%CITATION = MNRAA,371,1570;%%

%%\bibitem{Gruppuso2006}
%  A.~Gruppuso, C.~Burigana and F.~Finelli,
%  %``Dipole Straylight Contamination and Low Multipoles,''
%  PoS C {\bf MB2006}, 070 (2006)
%  [arXiv:astro-ph/0607413].
%  %%CITATION = POSCI,CMB2006,070;%%

%\bibitem{Efstathiou2003}
%  G.~Efstathiou,
%  %``Is the low CMB quadrupole a signature of spatial curvature?,''
%  Mon.\ Not.\ Roy.\ Astron.\ Soc.\  {\bf 343}, L95 (2003)
%  [arXiv:astro-ph/0303127].
%  %%CITATION = MNRAA,343,L95;%%

%\bibitem{Contaldi2003}
%  C.~R.~Contaldi, M.~Peloso, L.~Kofman and A.~Linde,
%  %``Suppressing the lower Multipoles in the CMB Anisotropies,''
%  JCAP {\bf 0307}, 002 (2003)
%  [arXiv:astro-ph/0303636].
%  %%CITATION = JCAPA,0307,002;%%

%\bibitem{Piao2003}
%  Y.~S.~Piao, B.~Feng and X.~m.~Zhang,
%  %``Suppressing CMB quadrupole with a bounce from contracting phase to
%  %inflation,''
%  Phys.\ Rev.\  D {\bf 69}, 103520 (2004)
%  [arXiv:hep-th/0310206].
%  %%CITATION = PHRVA,D69,103520;%%

%\bibitem{Kawasaki2003}
%  M.~Kawasaki and F.~Takahashi,
%  %``Inflation model with lower multipoles of the CMB suppressed,''
%  Phys.\ Lett.\  B {\bf 570}, 151 (2003)
%  [arXiv:hep-ph/0305319].
%  %%CITATION = PHLTA,B570,151;%%

%\bibitem{Tsujikawa2003}
%  S.~Tsujikawa, R.~Maartens and R.~Brandenberger,
%  %``Non-commutative inflation and the CMB,''
%  Phys.\ Lett.\  B {\bf 574}, 141 (2003)
%  [arXiv:astro-ph/0308169].
%  %%CITATION = PHLTA,B574,141;%%

%\bibitem{Moroi2003}
%  T.~Moroi and T.~Takahashi,
%  %``Correlated isocurvature fluctuation in quintessence and suppressed CMB
%  %anisotropies at low multipoles,''
%  Phys.\ Rev.\ Lett.\  {\bf 92}, 091301 (2004)
%  [arXiv:astro-ph/0308208].
%  %%CITATION = PRLTA,92,091301;%%

%\bibitem{Gordon2004}
%  C.~Gordon and W.~Hu,
%  %``A Low CMB Quadrupole from Dark Energy Isocurvature Perturbations,''
%  Phys.\ Rev.\  D {\bf 70}, 083003 (2004)
%  [arXiv:astro-ph/0406496].
%  %%CITATION = PHRVA,D70,083003;%%

%\bibitem{Weeks2003}
%  J.~Weeks, J.~P.~Luminet, A.~Riazuelo and R.~Lehoucq,
%  %``Well-proportioned universes suppress CMB quadrupole,''
%  Mon.\ Not.\ Roy.\ Astron.\ Soc.\  {\bf 352}, 258 (2004)
%  [arXiv:astro-ph/0312312].
%  %%CITATION = MNRAA,352,258;%%

%\bibitem{Piao2005}
%  Y.~S.~Piao,
%  %``A Possible Explanation to Low CMB Quadrupole,''
%  Phys.\ Rev.\  D {\bf 71}, 087301 (2005)
%  [arXiv:astro-ph/0502343].
%  %%CITATION = PHRVA,D71,087301;%%

%\bibitem{Wu2006}
%  C.~H.~Wu, K.~W.~Ng, W.~Lee, D.~S.~Lee and Y.~Y.~Charng,
%  %``Quantum noise and a low cosmic microwave background quadrupole,''
%  JCAP {\bf 0702}, 006 (2007)
%  [arXiv:astro-ph/0604292].
%  %%CITATION = JCAPA,0702,006;%%

%\bibitem{campanelli}
%  L.~Campanelli, P.~Cea and L.~Tedesco,
%  %``Ellipsoidal Universe Can Solve The CMB Quadrupole Problem,''
%  Phys.\ Rev.\ Lett.\  {\bf 97}, 131302 (2006)
%  [Erratum-ibid.\  {\bf 97}, 209903 (2006)]
%  [arXiv:astro-ph/0606266].
%  %%CITATION = PRLTA,97,131302;%%
%%\cite{Mather:1998gm}

%\bibitem{paoloceapaper}
%  P.~Cea,
%  %``Ellipsoidal Universe Induces Large Scale CMB Polarization,''
%  arXiv:astro-ph/0702293.
%  %%CITATION = ASTRO-PH/0702293;%%

%\bibitem{Mather}
%  J.~C.~Mather, D.~J.~Fixsen, R.~A.~Shafer, C.~Mosier and D.~T.~Wilkinson,
%  %``Calibrator Design for the COBE Far Infrared Absolute Spectrophotometer
%  %(FIRAS),''
%  Astrophys.\ J.\  {\bf 512} (1999) 511
%  [arXiv:astro-ph/9810373].
%  %%CITATION = ASJOA,512,511;%%

%\bibitem{paolocea} 
%P.~Cea, private communication.

\end{thebibliography}
\end{document}